\documentstyle[aps,prl,epsf,multicol]{revtex}
\newcommand{\hD}{\hat D}
\newcommand{\DPS}{\displaystyle}
\begin{document}

%%%%%%%%%%%%%%%%%%%%%%%%%%%%%%%%%%%%%%%%%%%%%%%%%%%%%%%%%%%%%%%%%%%%%%%%%%%%%
%\baselineskip 3pc
%\baselineskip 2pc
%%%%%%%%%%%%%%%%%%%%%%%%%%%%%%%%%%%%%%%%%%%%%%%%%%%%%%%%%%%%%%%%%%%%%%%%%%%%%

%\preprint{gr-qc/0011035}

\draft

%\begin{titlepage}
\title{
  Dynamics of a string coupled to gravitational waves II\\
  \smallskip 
  {\normalsize \it
     --- Perturbations propagate along an infinite Nambu-Goto string ---
   }
}
\author{
Kouji NAKAMURA\footnote{E-mail address : kouchan@phys-h.keio.ac.jp}, 
}
\address{
Department of Physics, Keio University,
Hiyoshi Yokohama, 223-8521, Japan
}
\author{
Hideki ISHIHARA\footnote{E-mail address : ishihara@th.phys.titech.ac.jp}
}
\address{
Tokyo Institute of Technology,
Oh-Okayama Meguro-ku, Tokyo 152-0033, Japan
}
%\date{November 8, 2000}
\date{\today}
\maketitle

\begin{abstract}
The perturbative modes propagating along an infinite string are 
investigated within the framework of the gauge invariant
perturbation formalism on a spacetime containing a
self-gravitating straight string with a finite thickness. 
These modes are not included in our previous analysis.
We reconstruct the perturbation formalism to discuss these modes
and solve the linearized Einstein equation within the first
order with respect to the string oscillation amplitude. 
In the thin string case, we show that the oscillations of an
infinite string must involve the propagation of cosmic string
traveling wave.
\end{abstract}
%\end{titlepage}

\begin{multicols}{2}

%%%%%%%%%%%%%%%%%%%%%%%%%%%%%%%%%%%%%%%%%%%%%%%%%%%%%%%%%%%%%%%%%%%%%
%\section{Introduction}
%\label{sec:introduction}
%%%%%%%%%%%%%%%%%%%%%%%%%%%%%%%%%%%%%%%%%%%%%%%%%%%%%%%%%%%%%%%%%%%%%

%%%%%%%%%%%%%%%%%%%%%%%%%%%%%%%%%%%%%%%%%%%%%%%%%%%%%%%%%%%%%%%%%%%
% Commented out by kouchan on Jan. 18
%%%%%%%%%%%%%%%%%%%%%%%%%%%%%%%%%%%%%%%%%%%%%%%%%%%%%%%%%%%%%%%%%%%
%These spatially one-dimensional objects present a particular
%interest, because it has been thought that the cosmic strings
%radiate gravitational waves by their rapid oscillations and lead
%to presently observable effects\cite{Vachaspati}.
%Detection of the gravitational waves from the strings, which are
%remnants of the early universe, might be a crucial evidence of
%the spontaneous symmetry breaking in the early universe. 
%Thus, it is important to study the precise dynamics of cosmic
%stings and their gravitational effects. 
%
%
%*****************************************************************
%
%
%
In the simplest case, the dynamics of cosmic strings, which are
topological defects associated with the symmetry breaking in
unified theories\cite{Vilenkin_Shellard}, is
idealized by the Nambu-Goto action. 
If the self-gravity of the string is ignored (test string case),
the Nambu-Goto action admits oscillatory solutions.   
Then it is considered that strings emit gravitational emission
by these oscillation and gradually lose their kinetic energy\cite{Vachaspati}.
%%%%%%%%%%%%%%%%%%%%%%%%%%%%%%%%%%%%%%%%%%%%%%%%%%%%%%%%%%%%%%%%%%%
% Newly added on Jan.18 by kouchan
%%%%%%%%%%%%%%%%%%%%%%%%%%%%%%%%%%%%%%%%%%%%%%%%%%%%%%%%%%%%%%%%%%%
The precise dynamics of strings is important for the estimation
of the energy of the gravitational waves from cosmic strings.

%*****************************************************************

To clarify the precise dynamics of them, we considered the
dynamics of an infinite self-gravitating Nambu-Goto string
within the first order with respect to its oscillation amplitude
in the previous paper\cite{kouchan-kneq0}.
%%%%%%%%%%%%%%%%%%%%%%%%%%%%%%%%%%%%%%%%%%%%%%%%%%%%%%%%%%%%%%%%%%%
% Commented out on Jan.18 by kouchan 
%%%%%%%%%%%%%%%%%%%%%%%%%%%%%%%%%%%%%%%%%%%%%%%%%%%%%%%%%%%%%%%%%%%
%Since there is no simple prescription of an arbitrary line source
%where a metric becomes singular\cite{Geroch}, we investigated
%first the dynamics of a thick Nambu-Goto string.
%After that, we consider its thin string situation in which the
%string thickness is much smaller than the wavelength of
%gravitational waves.
%We solved the scattering problem of gravitational waves by this
%infinite string.
%%%%%%%%%%%%%%%%%%%%%%%%%%%%%%%%%%%%%%%%%%%%%%%%%%%%%%%%%%%%%%%%%%%
% Modified on Jan.20 by kouchan 
%%%%%%%%%%%%%%%%%%%%%%%%%%%%%%%%%%%%%%%%%%%%%%%%%%%%%%%%%%%%%%%%%%%
%%%%%%%%% original %%%%%%%%%%%%%%%%%%%%%%%%%%%%%%%%%%%%%%%%%%%%%%%%
%It is shown that the string displacement is directly given by
%the variables of gravitational wave and there is no resonance in
%the scattering data for the thin string situation.
%%%%%%%%% new %%%%%%%%%%%%%%%%%%%%%%%%%%%%%%%%%%%%%%%%%%%%%%%%%%%%%
It is shown that the string displacement is directly determined by
the gravitational waves and there is no resonance in the
gravitational wave scattering by a thin string.
%%%%%%%%%%%%%%%%%%%%%%%%%%%%%%%%%%%%%%%%%%%%%%%%%%%%%%%%%%%%%%%%%%%
%%%%%%%%%%%%%%%%%%%%%%%%%%%%%%%%%%%%%%%%%%%%%%%%%%%%%%%%%%%%%%%%%%%
% Modified on Jan.20 by kouchan 
%%%%%%%%%%%%%%%%%%%%%%%%%%%%%%%%%%%%%%%%%%%%%%%%%%%%%%%%%%%%%%%%%%%
%%%%%%%%% original %%%%%%%%%%%%%%%%%%%%%%%%%%%%%%%%%%%%%%%%%%%%%%%%
%This shows that the dynamics of a test string and a
%self-gravitating one is quite different. 
%An infinite string bends only when the incidental wave is
%passing through the string 
%worldsheet and a string does not spontaneously oscillate without
%gravitational wave, while test strings can freely oscillate.
%%%%%%%%% new %%%%%%%%%%%%%%%%%%%%%%%%%%%%%%%%%%%%%%%%%%%%%%%%%%%%%
This shows an infinite string bends only when the gravitational
wave is passing through the string and a string does not
spontaneously oscillate without gravitational wave, in contrast
to the fact that test strings can freely oscillate. 
%%%%%%%%%%%%%%%%%%%%%%%%%%%%%%%%%%%%%%%%%%%%%%%%%%%%%%%%%%%%%%%%%%%

%*****************************************************************

Though the analyses in \cite{kouchan-kneq0} is important for the
precise dynamics of strings, these are insufficient to conclude
that there is no dynamical degree of freedom for free
oscillations of an infinite self-gravitating Nambu-Goto string.  
Because the perturbative modes which propagate along the string
with the light velocity is not included in the analyses in
\cite{kouchan-kneq0}.   
%%%%%%%%%%%%%%%%%%%%%%%%%%%%%%%%%%%%%%%%%%%%%%%%%%%%%%%%%%%%%%%%%%%
% Commented out by kouchan Jan.19
%%%%%%%%%%%%%%%%%%%%%%%%%%%%%%%%%%%%%%%%%%%%%%%%%%%%%%%%%%%%%%%%%%%
%because the vector and tensor 
%harmonics which are used in \cite{kouchan-kneq0} fail to be
%independent each other. 
%%%%%%%%%%%%%%%%%%%%%%%%%%%%%%%%%%%%%%%%%%%%%%%%%%%%%%%%%%%%%%%%%%%
% Newly added by kouchan Jan.18
%%%%%%%%%%%%%%%%%%%%%%%%%%%%%%%%%%%%%%%%%%%%%%%%%%%%%%%%%%%%%%%%%%%
Since these modes describe the dynamics of a test Nambu-Goto
string, one might expect that these modes contain the degree of
freedom for free oscillations of a self-gravitating infinite string. 
Thus, these modes are crucial for the dynamics of an infinite string.

%*****************************************************************

%In this article, we only consider these modes.
%%%%%%%%%%%%%%%%%%%%%%%%%%%%%%%%%%%%%%%%%%%%%%%%%%%%%%%%%%%%%%%%%%%
% Modified by kouchan Jan.19
%%%%%%%%%%%%%%%%%%%%%%%%%%%%%%%%%%%%%%%%%%%%%%%%%%%%%%%%%%%%%%%%%%%
%%%%%%%%% original %%%%%%%%%%%%%%%%%%%%%%%%%%%%%%%%%%%%%%%%%%%%%%%%
%We concentrate on the perturbative oscillations of an
%infinite string at the first order with respect to the
%oscillation amplitude of the string and show the oscillations of
%an infinite thin Nambu-Goto string are directly given by the
%propagation of the gravitational waves in these modes again.
%%%%%%%%% new %%%%%%%%%%%%%%%%%%%%%%%%%%%%%%%%%%%%%%%%%%%%%%%%%%%%%
In this article, we concentrate on these modes, which describe
the oscillations of an infinite string, within the first order
with respect to its oscillation amplitude.
We first solve the dynamics of an infinite thick string, after
that, we consider the thin string situation as in \cite{kouchan-kneq0}.
We show the oscillations of an infinite thin Nambu-Goto string
are directly given by the propagation of the gravitational waves.
%%%%%%%%%%%%%%%%%%%%%%%%%%%%%%%%%%%%%%%%%%%%%%%%%%%%%%%%%%%%%%%%%%%
These are just cosmic string traveling waves discovered by
Vachaspati\cite{traveling_wave-Vachaspati} and
Garfinkle\cite{traveling_wave-other}.  
%This is quite different from the behavior of an infinite
%test string; a test string can freely oscillate.

%%%%%%%%%%%%%%%%%%%%%%%%%%%%%%%%%%%%%%%%%%%%%%%%%%%%%%%%%%%%%%%%%%%%%
%\section{Background Spacetime}
%\label{wall_to_string}
%%%%%%%%%%%%%%%%%%%%%%%%%%%%%%%%%%%%%%%%%%%%%%%%%%%%%%%%%%%%%%%%%%%%%

As the background for the perturbation, we consider first a spacetime
$({\cal M},g_{\mu \nu})$ containing a straight thick
string\cite{kouchan-kneq0,stright_string}. 
The surface ${\cal S}$ of the thick string divides ${\cal M}$
into two regions: $ {\cal M}_{ex}$ and ${\cal M}_{in}$. 
${\cal M}_{in}$ is the `thick' world sheet of the string. 
%
%
%*****************************************************************
%
%
%%%%%%%%%%%%%%%%%%%%%%%%%%%%%%%%%%%%%%%%%%%%%%%%%%%%%%%%%%%%%%%%%%%
% Modified by kouchan Jan.19
%%%%%%%%%%%%%%%%%%%%%%%%%%%%%%%%%%%%%%%%%%%%%%%%%%%%%%%%%%%%%%%%%%%
%%%%%%%%% original %%%%%%%%%%%%%%%%%%%%%%%%%%%%%%%%%%%%%%%%%%%%%%%%
%We also divide ${\cal M}$ into two submanifolds so that 
%${\cal M} = {\cal M}_{1}\times{\cal M}_{2}$ and assume the
%background metric on ${\cal M}$ in the form
%\begin{equation}
%  \label{bg_metric}
%  ds^{2} = \gamma_{ab}dy^{a}dy^{b} + \eta_{pq}dz^{p}dz^{q},
%\end{equation}
%where $\gamma_{ab}$, the metric on ${\cal M}_{1}$, is given by 
%\begin{equation}
%  \label{homo_BG}
%  \gamma_{ab}dy^{a}dy^{b} 
%  = \frac{dr^{2}}{1 - \hat{\alpha}^{2}r^{2}} + r^{2}d\phi^{2}
%  = d\rho^2 + \frac{\sin^2\hat{\alpha}\rho}{\hat{\alpha}^2} d\phi^2, 
%  \quad \hat{\alpha}^{2} =  \frac{{\cal R}}{2} = 8 \pi G \sigma,
%\end{equation}
%on ${\cal M}_{in}\cap{\cal M}_{1}$,
%\begin{equation}
%  \label{vacuum_BG}
%  \gamma_{ab}dy^{a}dy^{b} =  \frac{dr^{2}}{(1 - \alpha)^{2}} +
%  r^{2}d\phi^{2} = d\rho^2 + (1 - \alpha)^2 \rho^2 d\phi^2, 
%\end{equation}
%on ${\cal M}_{ex}\cap{\cal M}_{1}$, and $\eta_{pq}$, the metric
%on ${\cal M}_{2}$, is the two dimensional Minkowski metric.
%%%%%%%%% new %%%%%%%%%%%%%%%%%%%%%%%%%%%%%%%%%%%%%%%%%%%%%%%%%%%%%
We also divide ${\cal M}$ into two submanifolds: %so that 
${\cal M} = {\cal M}_{1}\times{\cal M}_{2}$ and assume the
background metric on ${\cal M}$ in the form
\begin{equation}
  \label{bg_metric}
  ds^{2} = \eta_{pq}dz^{p}dz^{q} + \gamma_{ab}dy^{a}dy^{b}.
\end{equation}
The metric $\eta_{pq}$ on ${\cal M}_{2}$ is the two
dimensional Minkowski metric and the metric $\gamma_{ab}$ on
${\cal M}_{1}$ is given by  
\[
    \gamma_{ab}dy^{a}dy^{b} 
    = \frac{dr^{2}}{1 - \hat{\alpha}^{2}r^{2}} + r^{2}d\phi^{2}
    = d\rho^2 
    + \frac{\sin^2\hat{\alpha}\rho}{\hat{\alpha}^2} d\phi^2,
\]
on ${\cal M}_{in}\cap{\cal M}_{1}$, where $\hat{\alpha}^{2} =
{\cal R}/2 = 8 \pi G \sigma_{0}$ and $\phi\in[0,2\pi)$, and
\[
  \gamma_{ab}dy^{a}dy^{b} = \frac{dr^{2}}{(1-\alpha)^{2}} +
  r^{2}d\phi^{2} = d\rho^2 + (1 - \alpha)^2 \rho^2 d\phi^2, 
\]
on ${\cal M}_{ex}\cap{\cal M}_{1}$.
%%%%%%%%%%%%%%%%%%%%%%%%%%%%%%%%%%%%%%%%%%%%%%%%%%%%%%%%%%%%%%%%%%%
We shall use the indices $a,...,d$ for tensors on 
${\cal M}_{1}$ and $p,...,s$ for those on ${\cal M}_{2}$.
The Ricci scalar curvature ${\cal R}$ on ${\cal M}_{1}$ is
assumed to be a constant, and $\alpha$ is a deficit angle on 
${\cal M}_{ex}\cap{\cal M}_{1}$.   
The metric (\ref{bg_metric}) is a solution to the Einstein
equation with the energy momentum
tensor\cite{Vilenkin_Shellard,kouchan-kneq0,stright_string}: 
\begin{equation}
  \label{ene_mon}
  T_{\mu\nu} = - \sigma \eta_{\mu\nu}, 
  \quad \sigma = \left\{
    \begin{array}{ccc}
      \sigma_{0} &\mbox{for}& (r\le r_{*}), \\
      0 &\mbox{for}& (r > r_{*}), 
    \end{array}
  \right.
\end{equation}
where $\eta_{\mu\nu}$ is the four dimensional extension of
$\eta_{pq}$ and $r_{*}$ is the circumference radius of ${\cal
S}\cap{\cal M}_{1}$.
Israel's conditions \cite{Israel} at ${\cal S}$ yields 
$\alpha = 1 - \sqrt{1 - \hat{\alpha}^{2} r_{*}^{2}}$.

%*****************************************************************

%%%%%%%%%%%%%%%%%%%%%%%%%%%%%%%%%%%%%%%%%%%%%%%%%%%%%%%%%%%%%%%%%%%%%%
%\section{Perturbations}
%\label{perturbation}
%%%%%%%%%%%%%%%%%%%%%%%%%%%%%%%%%%%%%%%%%%%%%%%%%%%%%%%%%%%%%%%%%%%%%

%
%
%%%%%%%%%%%%%%%%%%%%%%%%%%%%%%%%%%%%%%%%%%%%%%%%%%%%%%%%%%%%%%%%%%%%%
%\subsection{Gauge Invariant Variables}
%%%%%%%%%%%%%%%%%%%%%%%%%%%%%%%%%%%%%%%%%%%%%%%%%%%%%%%%%%%%%%%%%%%%%
%
%
%
To consider the metric and matter perturbations on the
background (\ref{bg_metric}), we expand the perturbative
variables by the harmonics on ${\cal M}_{2}$.  
%%%%%%%%%%%%%%%%%%%%%%%%%%%%%%%%%%%%%%%%%%%%%%%%%%%%%%%%%%%%%%%%%%%
% Modified by kouchan Jan.19
%%%%%%%%%%%%%%%%%%%%%%%%%%%%%%%%%%%%%%%%%%%%%%%%%%%%%%%%%%%%%%%%%%%
%%%%%%%%% original %%%%%%%%%%%%%%%%%%%%%%%%%%%%%%%%%%%%%%%%%%%%%%%%
%Since we concentrate on the perturbations propagating along the
%string with the light velocity, we introduce the scalar
%harmonics
%\begin{equation}
%  \label{scalar-harmonics}
%  S:= e^{-i\omega (t+\epsilon z)},
%\end{equation}
%where $\epsilon = \pm 1$. 
%%%%%%%%% new %%%%%%%%%%%%%%%%%%%%%%%%%%%%%%%%%%%%%%%%%%%%%%%%%%%%%
Since we concentrate only on the modes propagating along the
string with the light velocity, we introduce
\begin{equation}
  \label{scalar-harmonics}
  S:= e^{-i\omega (t+\epsilon z)}, \quad \epsilon = \pm 1
\end{equation}
as the scalar harmonics.
$\epsilon$ determines the direction of the wave propagation.
%%%%%%%%%%%%%%%%%%%%%%%%%%%%%%%%%%%%%%%%%%%%%%%%%%%%%%%%%%%%%%%%%%%
For each $\omega$ and $\epsilon$, we introduce the null
vectors $k_{p}$ and $l_{p}$ defined by 
\begin{equation}
  \label{kp-kp-def}
%  \begin{array}{l}
    k_{p} S = - i \hD_{p}S, \quad %\\
    l_{p}l^{p} = 0, \quad l_{p}k^{p} = - 2\omega^{2}.
%  \end{array}
\end{equation}
Using these null vectors, tensor fields on ${\cal M}_{2}$ are
decomposed by the following harmonics:
\begin{equation}
  \label{vector-tensor-harmonics}
  \begin{array}{l}
  V_{(e1)}^{p} := i k^{p} S, \quad V_{(l1)}^{p} := i l^{p} S, \quad
  T_{(e0)pq} := \frac{1}{2}\eta_{pq} S, \\
  T_{(e2)pq} := - k_{p}k_{q} S, \quad
  T_{(l2)pq} := - l_{p}l_{q} S. 
  \end{array}
\end{equation}

%%%%%%%%%%%%%%%%%%%%%%%%%%%%%%%%%%%%%%%%%%%%%%%%%%%%%%%%%%%%%%%%%%%
% Separated the paragraph by kouchan Jan.19
%%%%%%%%%%%%%%%%%%%%%%%%%%%%%%%%%%%%%%%%%%%%%%%%%%%%%%%%%%%%%%%%%%%
%*****************************************************************

$S$, $V_{(e1)}^{p}$, $T_{(e0)pq}$ and $T_{(e2)pq}$ are same as 
those in \cite{kouchan-kneq0}. 
%%%%%%%%%%%%%%%%%%%%%%%%%%%%%%%%%%%%%%%%%%%%%%%%%%%%%%%%%%%%%%%%%%%
% Commented out by kouchan Jan.19
%%%%%%%%%%%%%%%%%%%%%%%%%%%%%%%%%%%%%%%%%%%%%%%%%%%%%%%%%%%%%%%%%%%
%, but $k_{p}$ in \cite{kouchan-kneq0} is not a null vector.
%%%%%%%%%%%%%%%%%%%%%%%%%%%%%%%%%%%%%%%%%%%%%%%%%%%%%%%%%%%%%%%%%%%
% Commented out by kouchan Jan.19
%%%%%%%%%%%%%%%%%%%%%%%%%%%%%%%%%%%%%%%%%%%%%%%%%%%%%%%%%%%%%%%%%%%
%%%%%%%%% original %%%%%%%%%%%%%%%%%%%%%%%%%%%%%%%%%%%%%%%%%%%%%%%%
%Since $k^{p}$ and $\epsilon^{pq}k_{p}$ ($\epsilon_{pq}$ is a
%two-dimensional antisymmetric tensor on ${\cal M}_{2}$) are
%linearly independent when $k^{p}$ is not null, then vector and
%tensor harmonics $V_{(o1)}^{p}=\epsilon^{pq}\hD_{q}S$ and 
%$T_{(o2)pq}=-\epsilon_{r(p}\hD_{q)}\hD^{r}S$ are used in
%\cite{kouchan-kneq0}. 
%%%%%%%%% new %%%%%%%%%%%%%%%%%%%%%%%%%%%%%%%%%%%%%%%%%%%%%%%%%%%%%
If $k^{p}$ is neither null nor zero, there are independent of
$V_{(o1)}^{p}:=\epsilon^{pq}\hD_{q}S$ and
$T_{(o2)pq}:=-\epsilon_{r(p}\hD_{q)}\hD^{r}S$ \cite{epsilonpq} 
in \cite{kouchan-kneq0}.
%%%%%%%%%%%%%%%%%%%%%%%%%%%%%%%%%%%%%%%%%%%%%%%%%%%%%%%%%%%%%%%%%%%
However, when $k^{p}$ is null, $V_{(o1)}^{p}$ ($T_{(o2)pq}$)
linearly depends on $V_{(e1)}^{p}$ ($T_{(e2)pq}$).
% as the vector (tensor) on ${\cal M}_{2}$ 
This is the reason why our analyses in \cite{kouchan-kneq0} fail
to include the modes propagating along the string with the
light velocity.
Instead, (\ref{scalar-harmonics}) and
(\ref{vector-tensor-harmonics}) are the set of independent
tensor harmonics for these modes and we expand the perturbation
variables by these harmonics.
%%%%%%%%%%%%%%%%%%%%%%%%%%%%%%%%%%%%%%%%%%%%%%%%%%%%%%%%%%%%%%%%%%%
% Commented out by kouchan Jan.19
%%%%%%%%%%%%%%%%%%%%%%%%%%%%%%%%%%%%%%%%%%%%%%%%%%%%%%%%%%%%%%%%%%%
%as shown in Fig.\ref{fig:fig1}.
%%%%%%%%%%%%%%%%%%%%%%%%%%%%%%%%%%%%%%%%%%%%%%%%%%%%%%%%%%%%%%%%%%%
% Added by kouchan Jan.19
%%%%%%%%%%%%%%%%%%%%%%%%%%%%%%%%%%%%%%%%%%%%%%%%%%%%%%%%%%%%%%%%%%%
%%%%%%%%%%%%%%%%%%%%%%%%%%%%%%%%%%%%%%%%%%%%%%%%%%%%%%%%%%%%%%%%%%%
%, and the one-to-one correspondence between the 
%perturbation of tensors of rank 2 and their Fourier components
%is guaranteed.  
%%%%%%%%%%%%%%%%%%%%%%%%%%%%%%%%%%%%%%%%%%%%%%%%%%%%%%%%%%%%%%%%%%%
% Commented out by kouchan Jan.19
%%%%%%%%%%%%%%%%%%%%%%%%%%%%%%%%%%%%%%%%%%%%%%%%%%%%%%%%%%%%%%%%%%%
%Thus, (\ref{scalar-harmonics}) and
%(\ref{vector-tensor-harmonics}) are appropriate to discuss the
%perturbations propagating along the string with the light
%velocity. 

%*****************************************************************

Let $h_{\mu\nu}$ be a perturbative metric and
$t^{\mu}_{\;\;\nu}$ be a perturbed energy-momentum tensor.
Using the harmonics (\ref{scalar-harmonics}) and
(\ref{vector-tensor-harmonics}), $h_{\mu\nu}$ and
$t^{\mu}_{\;\;\nu}$ are expanded as follows:  
\begin{eqnarray}
%  \label{metric_decomp_def_1}
  \nonumber
  &h_{ab} = {\displaystyle \int} f_{ab}S, \quad
  h_{ap} = {\displaystyle \int} \left\{f_{a(l1)}
  V_{(l1)p} + f_{a(e1)} V_{(e1)p}\right\},& \\
%  \label{metric_decomp_def_2}
  \nonumber
  &h_{pq} = {\displaystyle \int} \left\{f_{(l2)}
    T_{(l2)pq} + f_{(e0)} T_{(e0)pq}  + f_{(e2)} T_{(e2)pq}\right\},& \\
%  \label{e-m_decomp_def_1}
  \nonumber
  &t^{a}_{\;\;b} = {\displaystyle \int} s^{a}_{\;\;b}S, \quad
  t^{a}_{\;\;p} = {\displaystyle \int} \left\{ s^{a}_{(l1)} V_{(l1)p} 
    + s^{a}_{(e1)} V_{(e1)p}\right\},& \\
%  \label{e-m_decomp_def_2}
  \nonumber
  &t^{p}_{\;\;q} = {\displaystyle \int} \left\{s_{(l2)}
    {T_{(l2)}}^{p}_{\;\;q} + s_{(e0)} {T_{(e0)}}^{p}_{\;\;q} 
    + s_{(e2)} {T_{(e2)}}^{p}_{\;\;q}\right\},& 
\end{eqnarray}
where $\int := \sum_{\epsilon=\pm1}\int d\omega$.
The expansion coefficients are tensors on ${\cal M}_{1}$.
The perturbative energy momentum tensor $t^{\mu}_{\;\;\nu}$ has
its support only on ${\cal M}_{in}$.

%*****************************************************************

Here we consider the gauge-transformation of $h_{\mu\nu}$ and
$t^{\mu}_{\;\;\nu}$ associated with 
$x^{\mu} \rightarrow x^{\mu} + \xi^{\mu}$.
$\xi^{\mu}$ is expanded as 
\begin{equation}
  \label{xi_decomp_def}
  \xi_{a} := {\DPS \int} \zeta_{a}S, \quad
  \xi_{p} := {\DPS \int} \left\{\zeta_{(l1)} V_{(l1)p} 
    + \zeta_{(e1)} V_{(e1)p}\right\}. 
\end{equation}
Inspecting the gauge transformed variables 
$h_{\mu\nu} - {\pounds}_{\xi}g_{\mu\nu}$ and 
$t^{\mu}_{\;\;\nu} - {\pounds}_{\xi}T^{\mu}_{\;\;\nu}$, we find
simple gauge-invariant combinations: For the metric perturbations,
\begin{equation}
  \label{H_H_a_H_ab_def}
  \begin{array}{l}
    \DPS H := f_{(l2)}, \quad   
    H_{a} := f_{a(l1)} - \frac{1}{4\omega^{2}} D_{a} f_{(e0)}, \\
    H_{ab} := f_{ab} - 2 D_{(a}X_{b)}.
  \end{array}
\end{equation}
where $D_a$ is the covariant derivative associated with
$\gamma_{ab}$ and 
$X^{a} := f^{a}_{(e1)} - \frac{1}{2} D^{a}f_{(e2)}$ is
transformed to $X^{a} - \zeta^{a}$ by the gauge transformation. 
For the perturbations of $T^{\mu}_{\;\;\nu}$,
\begin{equation}
  \label{Sigma-V-def}
  \begin{array}{c}
    \Sigma := 16\pi G (s_{(e0)} + 2 X^{a}D_{a}\sigma), \\
    V^{a} := 16\pi G (s^{a}_{(e1)} - \sigma X^{a}). 
  \end{array}
\end{equation}
(\ref{Sigma-V-def}) are same as those of
$k^{p}k_{p}\neq 0$ modes in \cite{kouchan-kneq0} and all coefficients in
${t^{\mu}}_{\nu}$ except for $s_{(e0)}$ and $s^{a}_{(e1)}$ are
gauge invariant by themselves. 
$\Sigma$ is the perturbation of the string energy density which
is equal to the tangential tension.
$-V^{a}/{\cal R}$ is the displacement perturbation.

%*****************************************************************

%%%%%%%%%%%%%%%%%%%%%%%%%%%%%%%%%%%%%%%%%%%%%%%%%%%%%%%%%%%%%%%%%%%
% Modified by kouchan Jan.19
%%%%%%%%%%%%%%%%%%%%%%%%%%%%%%%%%%%%%%%%%%%%%%%%%%%%%%%%%%%%%%%%%%%
%%%%%%%%% original %%%%%%%%%%%%%%%%%%%%%%%%%%%%%%%%%%%%%%%%%%%%%%%%
%In this paper, we consider the perturbative motion of an
%infinite Nambu-Goto string within the first order with respect
%to its oscillation amplitude as discussed in \cite{kouchan-kneq0}. 
%We concentrate only on $\Sigma$ and $V^{a}$ and drop the other
%coefficients in the perturbative energy momentum tensor
%(\ref{e-m_decomp_def_1}) and (\ref{e-m_decomp_def_2}). 
%%%%%%%%% new %%%%%%%%%%%%%%%%%%%%%%%%%%%%%%%%%%%%%%%%%%%%%%%%%%%%%
In this paper, we consider the perturbative motion of an
infinite Nambu-Goto string.
% within the first order with respect
%to its oscillation amplitude.
We only consider $\Sigma$ and $V^{a}$ and drop the other
coefficients in the above perturbative energy momentum
tensor as discussed in \cite{kouchan-kneq0}.
In terms of the gauge invariant variables defined by
(\ref{H_H_a_H_ab_def}) and (\ref{Sigma-V-def}),
all components of the linearized Einstein equations on 
${\cal M}_{in}$ are given by 
\begin{equation}
\label{einstein-1}
  \begin{array}{l}
    \Delta H_{ab} = {\cal R} H_{ab} + 2 D_{(a}V_{b)} - 
    \gamma_{ab} D_{c}V^{c}, \\
    \Delta H_{a} - D^{c}D_{a}H_{c} - 2 \omega^{2} D_{a} H = 0, \\
    \Delta H = 0, \\
    D^{c}H_{ac} + 2 \omega^{2} H_{a} = V_{a}, \\
    D^{a}H_{a} + 2 \omega^{2} H = 0, \\
    H^{c}_{c} = 0,
  \end{array}
\end{equation}
where $\Delta := D^{a}D_{a}$. 
On ${\cal M}_{ex}$, the linearized Einstein equations are given
by setting ${\cal R} = 0 = V^{a}$ in (\ref{einstein-1}).
The perturbation of the divergence of $T_{\mu\nu}$ gives
\begin{eqnarray}
  \label{div_tmunu}
  2 \omega^{2} \sigma H^{a} = 0,  \quad
  D_{a}V^{a} + \frac{1}{2}\Sigma = 0. 
\end{eqnarray}
The first equation in (\ref{div_tmunu}) coincides with the
perturbation of the equation of motion derived from the Nambu-Goto
action\cite{kouchan-full}.  
The d'Alembertian of the string displacement in the first
equation vanishes by virtue of $k_{p}k^{p}=0$.
%%%%%%%%%%%%%%%%%%%%%%%%%%%%%%%%%%%%%%%%%%%%%%%%%%%%%%%%%%%%%%%%%%%

%*****************************************************************

The global solutions of the perturbations on ${\cal M}$ should
be constructed by matching the exterior and the interior
solutions to (\ref{einstein-1}) and (\ref{div_tmunu}) at ${\cal S}$.
This is accomplished by the perturbed Israel's junction conditions
$[\delta q_{\mu\nu}] := \delta q_{\mu\nu+} - \delta q_{\mu\nu-} = 0$ 
and $[\delta K^{\mu}_{\;\;\nu}] = 0$, where $\delta
q_{\mu\nu\pm}$ and $\delta K^{\mu}_{\;\;\nu\pm}$ are the
perturbed intrinsic metric and the extrinsic curvature of 
${\cal S}$, respectively.  
The subscripts $\pm$ represent the variables facing to 
${\cal M}_{ex}$ and ${\cal M}_{in}$, respectively.

%*****************************************************************

Choosing the gauge freedom $\zeta_{(l)\pm}$, $\zeta_{(e)\pm}$
and $\zeta^{a}_{\pm}$ defined by (\ref{xi_decomp_def}) so that
$f_{(e0)\pm}=f_{(e2)\pm}=0$ and  
$X^{a}_{+} = X^{a}_{-} = - V^{a}/{\cal R}$, 
the independent perturbative junction conditions for the 
perturbations at ${\cal S}$ are given by 
\begin{equation}
  \label{junction}
  \begin{array}{l}
    [H] = [n^{a}D_{a}H] = 0, \\
    \protect[n^{a}H_{a}] = [ \tau^{a}H_{a} ] = 0, \quad
    \protect[\epsilon^{cd}D_{c}H_{d}] = 0, \\
    \protect[\tau^{a}\tau^{a}H_{ab}] = [\tau^{a}n^{b}H_{ab}] = 0,
  \end{array}
\end{equation}
where $\epsilon^{cd}$ is a two-dimensional antisymmetric tensor
on ${\cal M}_{1}$ and
$n^{a}:=\left(\partial/\partial\rho\right)^{a}$ and
$\tau^{a}:=(1/r)\left(\partial/\partial\phi\right)^{a}$.

%*****************************************************************

We also impose that $H$, $n^{a}H_{a}$ and $\tau^{a}H_{a}$ are
finite at both $r\rightarrow\infty$ and $r=0$. 
Then, (\ref{einstein-1}), (\ref{div_tmunu}) and
(\ref{junction}) yield 
\begin{equation}
  H = 0 = H_{a}
\end{equation}
on the whole spacetime ${\cal M}$. 
Further, we find that the solution to
(\ref{einstein-1})-(\ref{div_tmunu}) are given by 
\begin{eqnarray}
    H_{ab} &=& D_{a}D_{b} \Phi_{out}, \nonumber\\
    \Phi_{out} &=& A_{0} \ln \rho + 
    \sum^{\infty}_{m=1}
    \left( A_{m} \rho^{\protect -\frac{\protect m}{\protect 1 - \alpha}}
      + B_{m}\rho^{\protect \frac{\protect m}{\protect 1 - \alpha}}
    \right) e^{im\phi}
  \label{outside_hab_sol_final}
\end{eqnarray}
on ${\cal M}_{ex}$, where $A_{m}$ and $B_{m}$ are arbitrary constants,
and 
\begin{eqnarray}
  && H_{ab} = \left(D_{a}D_{b} - \frac{1}{2}\gamma_{ab}\Delta\right)\Phi_{in}
  - \epsilon_{c(a}D_{b)}D^{c} \Psi_{in}, \nonumber\\
  \label{homo-sol-Hab}
%  \label{homo-sol-v-sigma}
  && V_{a} = \frac{1}{2} D_{a} (\Delta + {\cal R}) \Phi_{in} + \frac{1}{2}
  \epsilon_{ac} D^{c} (\Delta + {\cal R}) \Psi_{in}, \\
  && \Sigma = - \Delta (\Delta + {\cal R}) \Phi_{in}  \nonumber
\end{eqnarray}
on ${\cal M}_{in}$, without loss of generality\cite{kouchan-full}.
$V^{a}$ in (\ref{homo-sol-Hab}) shows that arbitrary $C^{4}$
functions $\Phi_{in}$ and $\Psi_{in}$ on ${\cal M}_{in}$ 
correspond to the irrotational and rotational part of the
internal matter velocity field of the thick string, respectively.

%*****************************************************************

(\ref{outside_hab_sol_final}) is just the cosmic string
traveling wave discovered by
Vachaspati\cite{traveling_wave-Vachaspati} and
Garfinkle\cite{traveling_wave-other}. 
This is the pp-wave exact solution to the vacuum
Einstein equation.
Imposing that $\tau^{a}\tau^{b}H_{ab}$ and $n^{a}\tau^{b}H_{ab}$
are finite at $r\rightarrow\infty$, we obtain $B_{m}=0$ for
$m\geq 1$ when $\alpha>1/2$, while only $B_{1}$ may not vanish
when $\alpha \leq 1/2$. 
The other coefficients in (\ref{outside_hab_sol_final}) 
depend on $\Phi_{in}$ and $\Psi_{in}$ by the conditions
(\ref{junction}).

%*****************************************************************

In this article, we consider the oscillatory behavior of a thin
string. 
``A thin string'' should be regarded as a string whose thickness
is much smaller than any other scales. 
The dynamics of a thin string is extracted from that of a thick
string by ignoring its internal fluctuations.
The thick string displacement is shown by the deformation
of ${\cal S}$, i.e., $X^{a}=-V^{a}/{\cal R}$ on ${\cal S}$.
$X^{a}$ itself represents various deformations of ${\cal S}$. 
Among them, the only $m=1$ mode in
(\ref{outside_hab_sol_final}), the dipole deformation,
represents the translation of ${\cal S}$ and it is relevant to
the string displacement as discussed in \cite{kouchan-kneq0}. 
Henceforce, we concentrate on the $m=1$ mode.

%*****************************************************************

%To see the behavior of a thin string, 
Here, we show examples of the global solutions under the
assumption that $\Phi_{in}$ and $\Psi_{in}$ are eigen functions
of $\Delta$ with a eigen value $\lambda$:  
\begin{equation}
  \label{eigenequation}
  \Delta \Phi_{in} = - \lambda \Phi_{in}, \quad 
  \Delta \Psi_{in} = - \lambda \Psi_{in}.
\end{equation}
Solutions on this assumption are sufficient to demonstrate
how a self-gravitating thin string behaves.

%*****************************************************************

When $\lambda\neq0$, the $m=1$ mode solutions to
(\ref{eigenequation}) are
\begin{equation}
  \label{inner-eigen-sol}
    \Phi_{in} = C_{1}P^{1}_{\nu}(x)e^{i\phi}, \quad
    \Psi_{in} = D_{1}P^{1}_{\nu}(x) e^{i\phi},
\end{equation}
where $x := \sqrt{1 - \hat{\alpha}^{2}r^{2}}$, 
$\nu(\nu + 1) = \lambda/\hat{\alpha}^2$.
When $\lambda=0$, we may choose $\Psi_{in}=0$ without loss of
generality and $\Phi_{in}$ of $m=1$ is given by 
\begin{equation}
  \label{lambda0-innersol}
  \Phi_{in} = \hat{C}_{1}
  \left(\frac{1-x}{1+x}\right)^{\frac{1}{2}} e^{i\phi}. 
\end{equation}
Here, $C_{1}$, $D_{1}$ and $\hat{C}_{1}$ are constants.
% which are 
%determined by the junction conditions (\ref{junction}).
To derive (\ref{inner-eigen-sol}) and (\ref{lambda0-innersol}), 
the regularity at $r=0$ is imposed.

%*****************************************************************

From (\ref{junction}), (\ref{outside_hab_sol_final}),
(\ref{homo-sol-Hab}) and (\ref{inner-eigen-sol}) or
(\ref{lambda0-innersol}), we obtain 
\begin{equation}
  \label{A_m-B_m-C_m-D_m-rel}
  \begin{array}{l}
    A_{1} = \frac{\displaystyle 1}{\displaystyle 2}
    \left(\frac{\displaystyle 1}{\displaystyle
      \rho_{*}}\right)^{-\frac{1}{1-\alpha}} 
    \left(C_{1}-i D_{1}\right) {\cal I}_{1}^{+}(\nu,x_{*}), \\
    B_{1} = \frac{\displaystyle 1}{\displaystyle 2}
    \left(\frac{\displaystyle 1}{\displaystyle
      \rho_{*}}\right)^{\frac{1}{1-\alpha}} 
    \left(C_{1} + i D_{1}\right) {\cal I}_{1}^{-}(\nu,x_{*})
  \end{array}
\end{equation}
for the $\lambda\neq 0$ case, and
\begin{equation}
  \label{lambda0-solution}
  A_{1}=0, \quad 
  B_{1} = \hat{C}_{1} \left(\frac{\alpha}{2-\alpha}\right)^{\frac{1}{2}}
  \rho_{*}^{- \frac{1}{1 - \alpha}} 
\end{equation}
for the $\lambda=0$ case, where 
$x_{*} := \sqrt{1 - \hat{\alpha}^{2}r^{2}_{*}}$ and  
\[
  \begin{array}{rcl}
    {\cal I}_{1}^{\pm}(\nu,x_{*}) := 
    &\pm&\sqrt{1 - x_{*}^{2}}P^{2}_{\nu}(x_{*}) \\
    &&+ 
  \frac{\DPS 1}{\DPS 2} (1 \mp x_{*})( 2 - \nu-\nu^{2}) 
  P^{1}_{\nu}(x_{*}). 
  \end{array}
\]

%*****************************************************************

The global solution (\ref{inner-eigen-sol}) and
(\ref{A_m-B_m-C_m-D_m-rel}) includes the situations in which the
string oscillates without gravitational waves outside the string.
When ${\cal I}^{\pm}_{1}=0$, both $A_{1}$ and $B_{1}$ may vanish 
while $C_{1}$ and $D_{1}$ are nonvanishing.
%%%%%%%%%%%%%%%%%%%%%%%%%%%%%%%%%%%%%%%%%%%%%%%%%%%%%%%%%%%%%%%%%%%
% Commented out by kouchan Jan.20
%%%%%%%%%%%%%%%%%%%%%%%%%%%%%%%%%%%%%%%%%%%%%%%%%%%%%%%%%%%%%%%%%%%
%\begin{equation}
%  \label{no-wave}
%  C_{1}=\pm iD_{1}\neq 0, \quad {\cal I}^{\mp}_{1}(\nu,x_{*})=0.
%\end{equation}
%%%%%%%%%%%%%%%%%%%%%%%%%%%%%%%%%%%%%%%%%%%%%%%%%%%%%%%%%%%%%%%%%%%
Since $V_{a}$ does not vanish in these case, these do
describe the oscillations of the string.
The equation ${\cal I}^{\pm}_{1}=0$ requires the relation 
between $\nu$ and $x_{*}$.
It can be shown that there is no solution in $0<\nu<1$
but the solutions exist if $\nu\geq 1$.

%*****************************************************************
%%%%%%%%%%%%%%%%%%%%%%%%%%%%%%%%%%%%%%%%%%%%%%%%%%%%%%%%%%%%%%%%%%%%%
%\section{Thick and Thin String}
%\label{thin string case}
%%%%%%%%%%%%%%%%%%%%%%%%%%%%%%%%%%%%%%%%%%%%%%%%%%%%%%%%%%%%%%%%%%%%%

However, these string oscillations without gravitational waves
are in the thick string case. 
In the global solutions given by (\ref{A_m-B_m-C_m-D_m-rel}) or 
(\ref{lambda0-solution}), the scale of the string thickness is
given by the curvature radius $\sim {\cal R}^{-1/2}$ on 
${\cal M}_{in}\cap{\cal M}_{1}$ with the fixed deficit angle
$\alpha$, while the scale of the internal fluctuations is 
given by $\lambda^{-1/2}$.
The condition $\lambda/{\cal R}\sim\nu\geq 1$ for the existence
of the solution to ${\cal I}^{\pm}_{1}=0$ corresponds to the
situation in which the scale of the internal fluctuations is
comparable to the string thickness.  
The string can oscillate without the gravitational wave only
when such small scale internal fluctuations are allowed.

%*****************************************************************

In contrast to the thick string case, ``a thin string case'' is
regarded as the situation $\nu\ll 1$ with the fixed $\alpha$. 
As the thin string case, we take the leading order of $\nu$ in
the global solutions.
When $\nu\rightarrow0$, ${\cal I}^{\pm}_{1}(\nu,x_{*})$ behave  
\begin{equation}
  \label{cal_I_pm_asympt}
  \begin{array}{l}
    {\cal I}^{+}_{1}(\nu,x_{*})= O(\nu^{2}), \\
    {\cal I}^{-}_{1}(\nu,x_{*})= 2\nu 
    \left(\frac{\DPS 1-x_{*}}{\DPS 1+x_{*}}\right)^{\frac{1}{2}} + O(\nu^{2}).
  \end{array}
\end{equation}
and all solutions (\ref{inner-eigen-sol}) with
(\ref{A_m-B_m-C_m-D_m-rel}) are reduced to a solution  
(\ref{lambda0-innersol}) with (\ref{lambda0-solution}) by the
replacement $\hat{C}_{1} = \nu C_{1}$.  
Thus, (\ref{lambda0-innersol}) with (\ref{lambda0-solution}) is
the unique solution describes the thin string oscillations.  
This also shows that the traveling wave $B_{1}$ in
(\ref{outside_hab_sol_final}) is not concerned with the internal
fluctuation, while the traveling wave $A_{m}$ in are excited by
the internal fluctuations of the matter field.

%*****************************************************************

In the same order of $\nu$, the string displacement is
\begin{eqnarray}
  \label{string_deformation-k0}
    X^{a}_{S} &:=& \DPS\int \left.X^{a} \right|_{{\cal
    S},m=1,\lambda = 0} S 
    = - \frac{1}{2} \int \left.S D^{a}\Phi_{out}
    \right|_{{\cal S},m=1,\lambda=0} \nonumber\\
  &=& - \DPS\int \frac{B_{1}}{2(1 - \alpha)} 
  \left(\frac{r_{*}}{1 - \alpha}\right)^{\frac{\alpha}{1-\alpha}} S
   e^{i\phi} \left( n^{a} + i \tau^{a} \right).
\end{eqnarray}
Since $X^{a}_{S}$ satisfy the equation of motion
(\ref{div_tmunu}) derived from the Nambu-Goto action,
$X^{a}_{S}$ is regarded as the displacement of 
the thin Nambu-Goto string.
$X^{a}_{S}$ in (\ref{string_deformation-k0}) is completely
determined by the cosmic string traveling wave $B_{1}$. 
%There is no dynamical degree of freedom concerning the
%spontaneous oscillations of an infinite Nambu-Goto string.
The oscillations of a string are nothing but the propagation of
the gravitational waves along it. 
This is our conclusion of the $k^{p}k_{p}=0$ modes analyses.

%%%%%%%%%%%%%%%%%%%%%%%%%%%%%%%%%%%%%%%%%%%%%%%%%%%%%%%%%%%%%%%%%%%%%
%\section{Summary and Discussion}
%%%%%%%%%%%%%%%%%%%%%%%%%%%%%%%%%%%%%%%%%%%%%%%%%%%%%%%%%%%%%%%%%%%%%

%*****************************************************************

The amplitude $|X^{a}_{S}|:=\sqrt{\gamma_{ab}X^{a}_{S}X^{b}_{S}}$ 
is estimated as follows:
Choosing the gauge freedom $\zeta_{(e)-}$ so that 
$f_{(e2)} = \Phi_{in}$, we obtain 
$h_{\mu\nu}dx^{\mu}dx^{\nu}=\int\Phi_{in}S\omega^2(dt + \epsilon dz)^2$. 
This shows the linear perturbation is not violated if
$|\Phi_{in}\omega^{2}|<1$.  
Since $|X^{a}_{S}|/r_{*} \sim |\Phi_{in}|\omega^2/(\omega r_{*})^2$, 
$|X^{a}_{S}|>r_{*}$ if $\omega r_{*}\ll 1$.
Therefore, the magnitude of the displacement may become larger
than the string thickness within the linear perturbation
framework if the wavelength of the traveling wave is
sufficiently larger than the string thickness.

%*****************************************************************

Though the global solutions discussed here are specific one,
our conclusion in the thin string case will be hold in the wider
class of the solutions (\ref{homo-sol-Hab}).
Since the eigen functions of $\Delta$ with different positive
eigen values span the $L^{2}$ space,
the internal fluctuations $\Phi_{in}$ and $\Psi_{in}$ in $L^{2}$
space are decomposed by these eigen functions.
Even so, in the thin string case, we neglect the fluctuations
described by the eigen functions whose eigen values are
comparable to or larger than the internal curvature.
Then we will obtain (\ref{lambda0-innersol}) as the solution
for the thin string case and will reach to the same conclusion as
here.

%*****************************************************************

We have seen that our conclusion in \cite{kouchan-kneq0}, which
states that there is no dynamical degree of freedom of the free
oscillations of an infinite Nambu-Goto string, is unchanged even
if the $k^{p}k_{p}=0$ modes are included into our consideration.
Our formalism in \cite{kouchan-kneq0} and here do not include
the cylindrical static perturbations, which would be 
necessary for completion, but these have nothing to do with the
dynamics of the string. 
Then, as the result of the systematic search of the dynamical
perturbations, we conclude that the small amplitude oscillations
of an infinite self-gravitating Nambu-Goto string are
gravitational wave propagation.
This is the essentially same conclusion as that for the
thin spherical wall case in \cite{Kodama}.
An infinite Nambu-Goto string bends only when gravitational
waves are passing through the string worldsheet and continues
to oscillate only when gravitational waves are propagating along
itself.

%*****************************************************************

The continuous string oscillations are the propagation of cosmic
string traveling waves. 
As pointed out by Vachaspati\cite{traveling_wave-Vachaspati},
the traveling wave does not emit gravitational waves in the
direction perpendicular to the string by itself.
If the second or higher order in the string's oscillation
amplitude give rise to the mode coupling of $k^{p}k_{p}=0$ and
$k^{p}k_{p}\neq 0$ modes, the string oscillations might emit the 
gravitational waves towards an observer far from the string.
In the thin wall case, if there exists the gravitational wave
solution propagating along the wall, the situation may be analogous.

%*****************************************************************

%%%%%%%%%%%%%%%%%%%%%%%%%%%%%%%%%%%%%%%%%%%%%%%%%%%%%%%%%%%%%%%%%%%
% Commented out by kouchan Jan.20
%%%%%%%%%%%%%%%%%%%%%%%%%%%%%%%%%%%%%%%%%%%%%%%%%%%%%%%%%%%%%%%%%%%
%We also note that a ``thin string'' in this paper is not
%a string with the zero thickness ($r_{*}=0$) as in
%\cite{kouchan-kneq0}. 
%The magnitude of the string displacement $X^{a}_{S}$ given in
%(\ref{string_deformation-k0}) depends on
%$r_{*}$ and vanishes in the limit $r_{*}\rightarrow 0$ with the
%fixed all other parameters; the deficit angle $\alpha$, the
%amplitude $B_{1}$ and the frequency of the gravitational wave
%are fixed. 
%This is same as our result in\cite{kouchan-kneq0} and 
%consistent with that obtained by Unruh et.al\cite{Unruh}. 
%This suggests that the thickness must remain finite when
%one investigates the dynamics of self-gravitating strings and
%their gravitational wave emission. 
%The cosmic strings formed during phase transition in the early
%universe have their thickness determined by the energy scale of
%the spontaneously symmetry breaking\cite{Vilenkin_Shellard}.
%However, since the angular deficit of the string is extremely
%small ($10^{-6}$ radian for GUT strings), the thickness dependence of
%the oscillation amplitude of strings will be extremely small.

%*****************************************************************

%\section*{acknowledgements}

We are pleased to thank Professor Minoru Omote and Professor
Akio Hosoya for their continuous encouragement. We wish also to
thank Professor Hideo Kodama for the valuable comments and
discussion.

%%%%%%%%%%%%%%%%%%%%%%%%%%%%%%%%%%%%%%%%%%%%%%%%%%%%%%%%%%%%%%%%%%%%%

\end{multicols}

%\begin{figure}[htbp]
%  \begin{center}
%    \leavevmode
%    \epsfxsize=0.8\textwidth
%    \epsfbox[0 0 1155 383]{fig1.eps}
%    \caption{$V_{(o1)}^{p}$ fails to be independent of
%    $V_{(e1)}^{p}$ when $k^{p}k_{p}=0$ and $V_{(l1)}^{p}$ is
%    introduced as an independent vector harmonics.}
%    \label{fig:fig1}
%  \end{center}
%\end{figure}


\begin{thebibliography}{99}
\bibitem{epsilonpq} 
  $\epsilon_{pq}$ is a two-dimensional antisymmetric tensor on
  ${\cal M}_{2}$.
\end{thebibliography}
\end{document}